\documentclass[twocolumn,shortnote]{jpsj3}

\usepackage{graphicx}% Include figure files
\usepackage{dcolumn}% Align table columns on decimal point
\usepackage{bm}% bold math
\usepackage{color}
\usepackage{ulem}

\begin{document}

\title{Highly Uniform Magnetic and Electronic Environment in Non-Centrosymmetric Superconductor LaRhGe$_3$}

\author{Shunsaku~Kitagawa$^{1,}$\thanks{E-mail address: kitagawa.shunsaku.8u@kyoto-u.ac.jp}, 
Hiroyasu Matsudaira$^{1}$, 
Kenji~Ishida$^{1}$, and
Mohamed Oudah$^{2}$
}

\inst{$^1$Department of Physics, Graduate School of Science, Kyoto University, Kyoto 606-8502, Japan \\
$^2$Stewart Blusson Quantum Matter Institute, University of British Columbia, Vancouver, BC, Canada
}

\date{\today}

\abst{
We report the results of $^{139}$La NMR measurements in the non-centrosymmetric superconductor LaRhGe$_3$. 
This material crystallizes in a tetragonal structure without inversion symmetry and exhibits type-I superconductivity below 385 mK.
We observed remarkably sharp NMR signals, indicating that the magnetic and electronic properties of the sample are extremely uniform in LaRhGe$_3$ despite the complex crystal structure.
Our NMR results indicate that LaRhGe$_3$ is a weakly correlated semimetal in the normal state.
}

\maketitle

%\section{Introduction}
The discovery of superconductivity in non-centrosymmetric systems has attracted significant attention due to the possibility of unconventional pairing states\cite{E.Bauer_PRL_2004,H.Takahashi_JPSJ_2023,A.Maurya_JPSJ_2023,N.Furutani_JPSJ_2023}. 
In centrosymmetric superconductors, the pairing interaction typically leads to a spin-singlet or spin-triplet state\cite{H.Tou_JPSJ_2005,S.Kitagawa_PRL_2023,H.Matsumura_JPSJ_2023,S.Ogawa2023,S.Kitagawa_JPSJ_2024}, but in non-centrosymmetric superconductors, the lack of inversion symmetry allows for an admixture of the spin-singlet and spin-triplet pairing states\cite{Gorkov_PRL_2001}. 
This opens the possibility of novel superconducting properties, which are not observed in conventional superconductors.

\begin{figure}
    \centering
    \includegraphics[width=\linewidth]{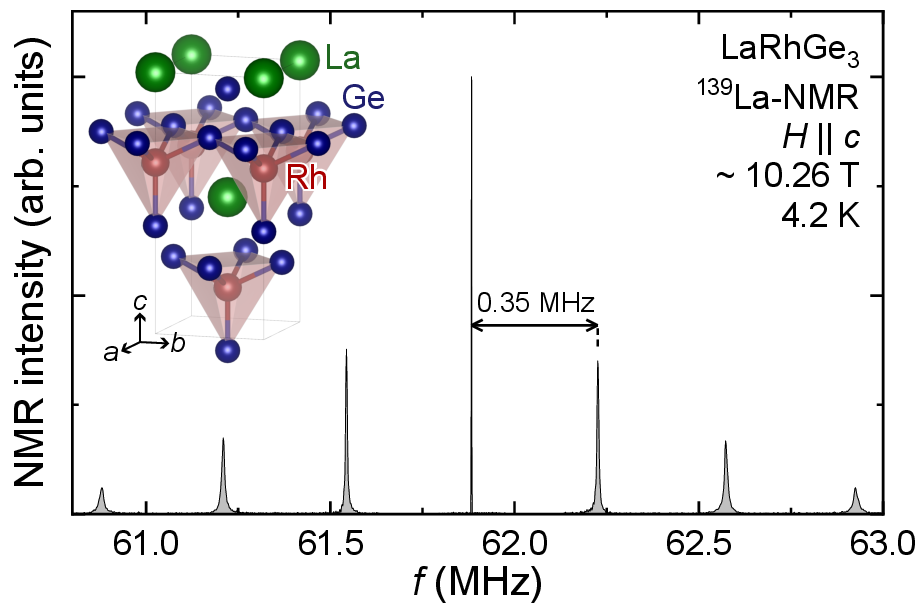}
    \caption{(Color online) NMR spectrum of $^{139}$La in LaRhGe$_3$ measured at 4.2 K. Seven sharp peaks are observed, which are attributed to quadrupole splitting due to the electric field gradient at the La site.
    (Inset) The crystal structure of LaRhGe$_3$ drawn by VESTA\cite{K.Momma_JAC_2011}.}
    \label{fig:NMR_spectrum}
\end{figure}

LaRhGe$_3$ is a non-centrosymmetric superconductor that crystallizes in a tetragonal structure with space group I4$mm$ ($C_{4v}^{9}$, \#107) (see the inset of Fig.\ref{fig:NMR_spectrum})\cite{Venturini1985} and exhibits type-I superconductivity with a critical temperature of 385 mK \cite{M.Oudah_npj_2024}.
The observation of a clear Meissner signal and a low critical magnetic field suggests a clean and homogeneous superconducting phase. 
However, the microscopic nature of the pairing state in LaRhGe$_3$ remains unresolved, particularly the possible admixture of the spin-singlet and spin-triplet components.

In addition to its superconducting properties, LaRhGe$_3$ has drawn attention due to its electron-phonon drag effects\cite{M.Oudah_npj_2024}.
The electron-phonon drag refers to a phenomenon where the phonon system contributes to the charge and heat transports through coupling with the electron system, which is realized in some clean systems\cite{H.-Y.Yang_NC_2021,G.B.Osterhoudt_PRX_2021}.
In the normal state of LaRhGe$_3$, the temperature dependence of electric resistivity at zero magnetic field deviates from a conventional Fermi liquid behavior and is well fitted by the electron-phonon drag model\cite{H.-Y.Yang_NC_2021}.
A peak observed around 30 K in the derivative of the resistivity, the Seebeck coefficient, and the thermal conductivity is consistent with the presence of the electron-phonon drag in this material.

In this study, we have performed $^{139}$La nuclear magnetic resonance(NMR) measurements to investigate the normal state of LaRhGe$_3$.
Notably, we observed sharp NMR signals, which indicate that the magnetic and electronic properties of the sample are extremely uniform.
Unfortunately, we could not detect the La nuclear quadrupole resonance (NQR) signals at zero magnetic field due to the limitation of our NMR/NQR spectrometer.

%\section{Experimental Methods}
High-quality single crystals of LaRhGe$_3$ were synthesized using the metallic self-flux method\cite{M.Oudah_npj_2024}. 
The frequency-swept $^{139}$La(nuclear spin $I$ = 3/2, nuclear gyromagnetic ratio $^{139}\gamma_{\rm n}/2\pi = 6.0142$~MHz/T, and natural abundance 99.91\%)-NMR spectra were acquired by performing the Fourier transform of the spin-echo signal observed after a spin-echo radiofrequency pulse sequence in a fixed magnetic field ($\sim$ 10~T) parallel to the $c$-axis.
The magnetic field was calibrated using a $^{63}$Cu ($^{63}\gamma_{\rm n}/2\pi = 11.285$ MHz/T)-NMR signal with the Knight shift $K_{\rm Cu} = 0.2385$\% from the NMR coil\cite{MetallicShift}.
Nuclear spin-lattice relaxation rate 1/$T_1$ was measured using the saturation-recovery method.
1/$T_1$ was determined by fitting a theoretical function for $I$ = 7/2.
The single-component $T_1$ was evaluated in the whole temperature range.

%\section{Results and Discussion}
Figure \ref{fig:NMR_spectrum} shows the $^{139}$La NMR spectrum of LaRhGe$_3$ at 4.2~K.
Seven extremely sharp NMR signals were observed.
Since the nuclear spin of $^{139}$La is 7/2, in the presence of an electric field gradient (EFG) at the La site, electric quadrupole interactions cause the NMR signals to split into seven peaks. 
The total Hamiltonian, including both the Zeeman interaction and the quadrupole interaction, is given by
\begin{align}
\mathcal{H} = - &\gamma \hbar (1 + K)I\cdot H + \notag\\
&\frac{h\nu_{\mathrm{Q}}}{6}\left[3I_{z}^{2}-I\left(I+1\right)+\frac{\eta}{2}\left(I_{+}^{2}+I_{-}^{2}\right)\right],
\label{eq.1}
\end{align}
where $K$ is the Knight shift, $h$ is the Planck constant, $\nu_{\mathrm{Q}} = \frac{3heQV_{zz}}{2I(2I-1)}$ is the NQR frequency, and $\eta = \left|\frac{V_{yy}-V_{xx}}{V_{zz}}\right|$ is the asymmetry parameter.
In LaRhGe$_3$, the parameter $\eta$ is zero at the La site due to the four-fold symmetry at the atomic position.
The NMR linewidth broadens due to inhomogeneous magnetic susceptibility and/or distribution in the EFG.
However, the full width at half maximum in the NMR spectrum of LaRhGe$_3$ is remarkably small, with 1 kHz for the center peak and 5 kHz for the first satellite peaks.
This is highly unusual for a complex crystal structure with the broken inversion symmetry\cite{PhysRevB.85.052501}.
The sharp linewidth indicates uniform magnetic and electronic properties throughout the sample, which is consistent with both the type-I superconductivity and the electron-phonon drag effects observed in LaRhGe$_3$, as both phenomena require high sample quality characterized by a long mean free path and weak electron-electron correlations.

From the NMR spectrum in Fig.~\ref{fig:NMR_spectrum}, the quadrupole frequency $\nu_{\mathrm{Q}}$ is estimated to be 0.35~MHz.
As LaRhGe$_3$ is a type-I superconductor, zero-field measurements are necessary to investigate the superconducting properties.
However, the La NQR spectrum in LaRhGe$_3$ has not been detected so far.
This raises open questions regarding the nature of the superconductivity in this material.

\begin{figure}[bt]
    \centering
    \includegraphics[width=\linewidth]{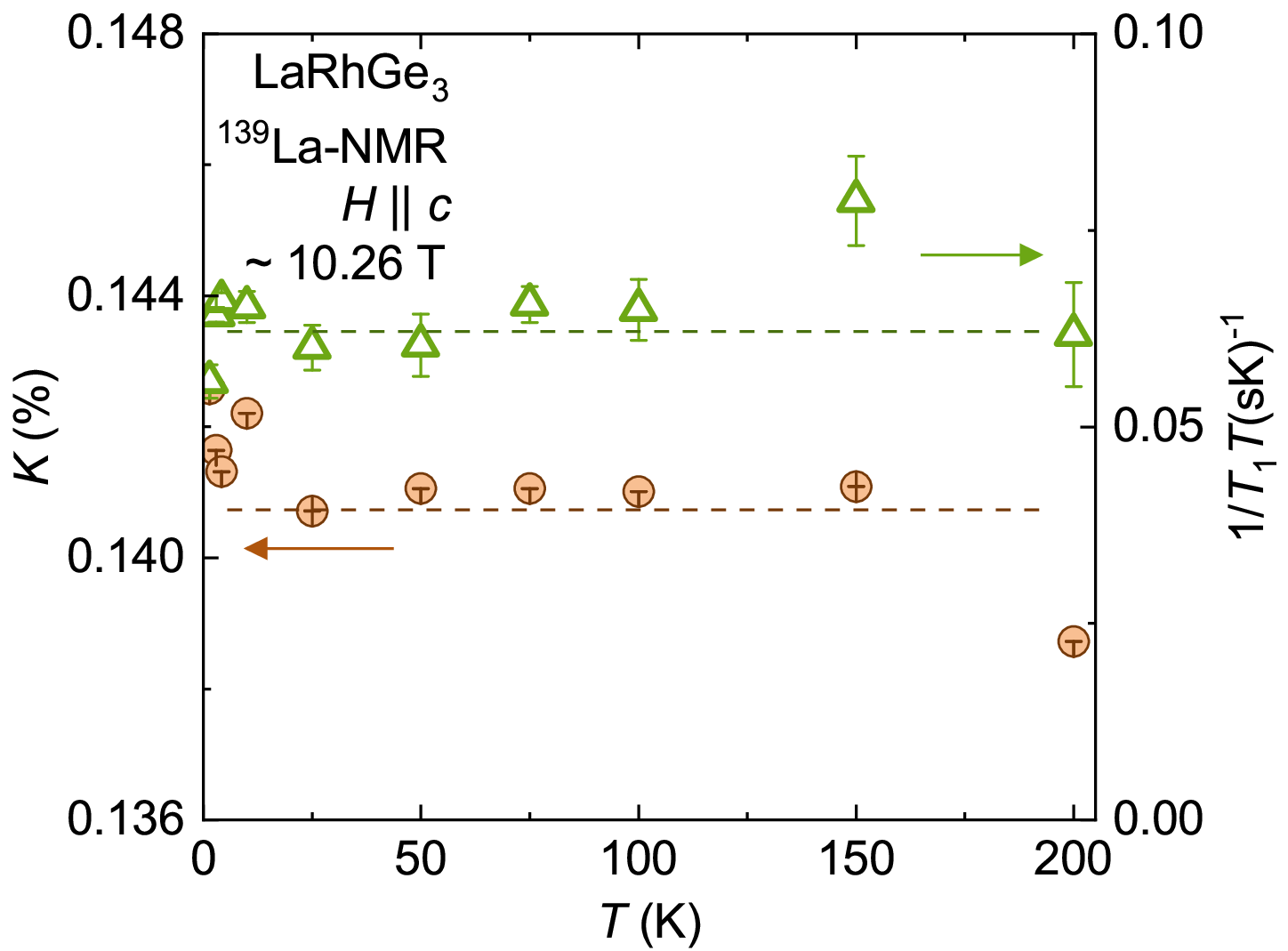}
    \caption{(Color online) Temperature dependence of the Knight shift (closed circles) and $1/T_1T$ (open triangles) in LaRhGe$_3$.
    The broken lines are guides for eyes.}
    \label{fig:Knight_shift}
\end{figure}

For the case where the magnetic field is applied parallel to the $c$-axis ($H \parallel c$), the center peak is not shifted by quadrupole interactions.
Therefore, using the resonance frequency of the center peak $f_{\mathrm{res}}$, the Knight shift $K$ can be determined as
\begin{align}
K (\%) = \frac{f_{\mathrm{res}} - f_0}{f_0} \times 100,
\end{align}
where $f_0 = (^{139}\gamma_{\rm n}/2\pi)\mu_0H$ is the Larmor frequency.
Figure \ref{fig:Knight_shift} shows the temperature dependence of the Knight shift and the spin-lattice relaxation rate divided by temperature, $1/T_1T$.
Both the Knight shift and $1/T_1T$ show constant over the all measured temperature range, indicating typical metallic behavior.
The small values of $K$ and 1/$T_1T$ indicate the small density of states, which is consistent with semimetallic behavior in the previous report\cite{M.Oudah_npj_2024}.
To further characterize the strength of electron correlations, we evaluated the Korringa ratio, $K(\alpha) = \frac{\hbar}{4 \pi k_{\rm B}} \left( \frac{\gamma_e}{^{139}\gamma_{\rm n}} \right)^2 \frac{1}{T_1 T K^2}$, where $k_{\rm B}$ is the Boltzmann constant, and $\gamma_e$ is the gyromagnetic ratio of the electron\cite{T.Moriya_JPSJ_1963}.
$K(\alpha)$ in LaRhGe$_3$ was estimated to be approximately 4, indicating weak antiferromagnetic correlations.

%\section{Conclusion}
In conclusion, we performed $^{139}$La NMR measurements to investigate the normal-state properties of the non-centrosymmetric superconductor LaRhGe$_3$.
A remarkably sharp NMR spectrum was observed.
The narrow linewidths of both the center and satellite peaks indicate an exceptionally uniform magnetic and electronic environment in LaRhGe$_3$, which is consistent with the type-I superconductivity and electron-phonon drag effects.
This is unusual for non-centrosymmetric ternary compounds.
To investigate the superconducting properties, the detection of the La NQR spectrum is desired.

\section*{acknowledgments}
We thank Samikshya Sahu and Alannah M. Hallas for their help with synthesis.
This work was supported by Grants-in-Aid for Scientific Research (KAKENHI Grant No. JP20KK0061, No. JP20H00130, No. JP21K18600, No. JP22H04933, No. JP22H01168, No. JP23H01124, No. JP23K22439 and No. JP23K25821) from the Japan Society for the Promotion of Science, by JST SPRING(Grant No. JPMJSP2110) from Japan Science and Technology Agency, by research support funding from The Kyoto University Foundation, by ISHIZUE 2024 of Kyoto University Research Development Program, by Murata Science and Education Foundation, and by the JGC-S Scholarship Foundation.
In addition, liquid helium is supplied by the Low Temperature and Materials Sciences Division, Agency for Health, Safety and Environment, Kyoto University.

\end{document}